\documentclass{aa}

\usepackage{graphicx}
\usepackage[varg]{txfonts}
\usepackage{natbib}
\usepackage[pdftex,colorlinks=true,linkcolor=blue,citecolor=blue,urlcolor=blue]{hyperref}

\def\i{\mathrm{i}}

\def\expo#1{\mathrm{e}^{#1}}
\def\d{\mathrm{d}}
\def\O#1{\mathrm{O}\left(#1\right)}

\graphicspath{{./figures/}}

\newcommand\figI{
  \begin{figure*}
    \centering
    \includegraphics[width=\linewidth]{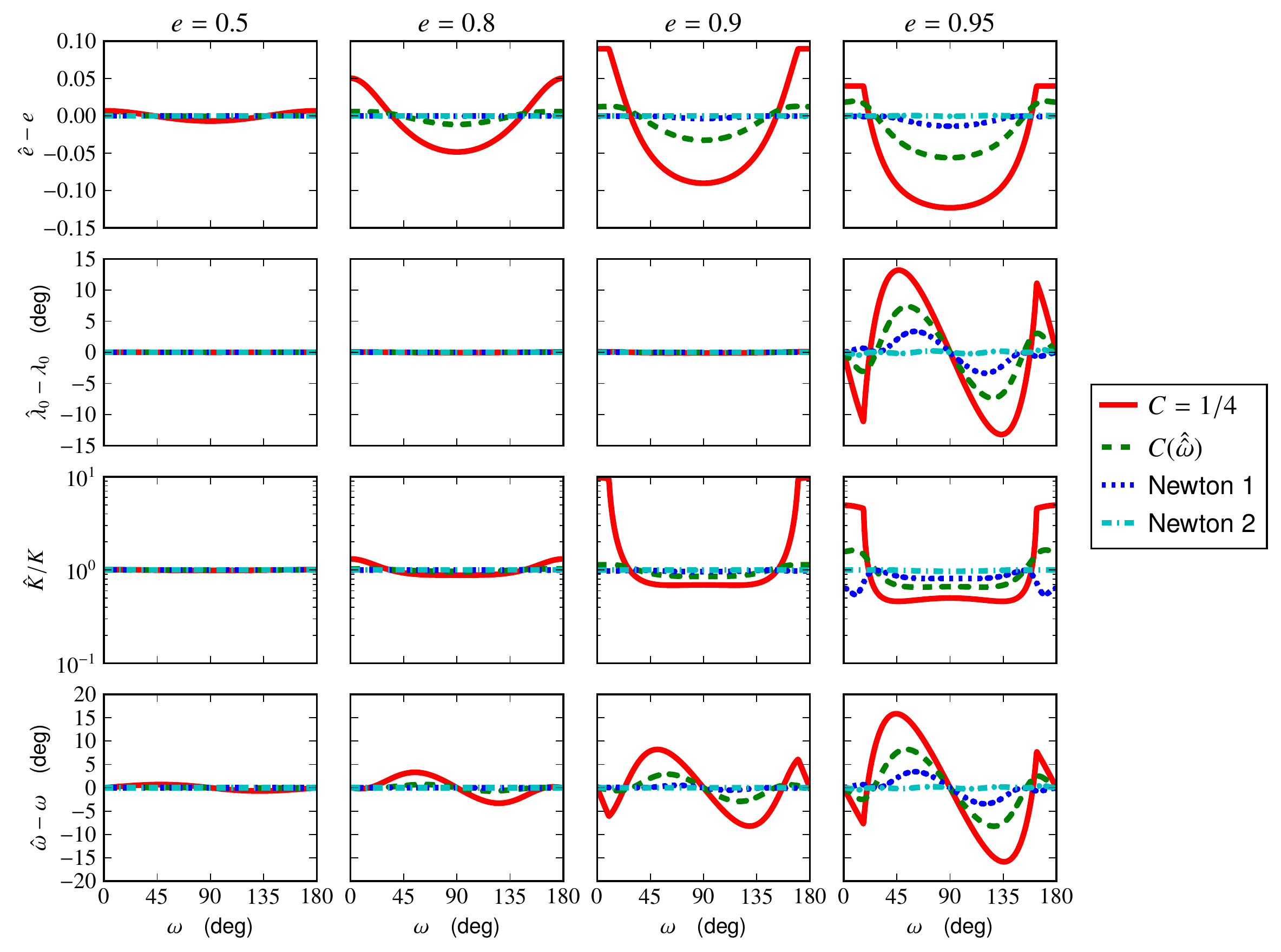}
    \caption{Comparison between our analytical estimates
      (see Eqs.~(\ref{eq:ecc}),~(\ref{eq:M0}),~(\ref{eq:C}),~(\ref{eq:S}))
      and the expected values of the eccentricity ($e$),
      the mean longitude at the reference time ($\lambda=M_0+\omega$),
      the semi-amplitude $K$,
      and the argument of periastron ($\omega$) (from \textit{top} to \textit{bottom}),
      as a function of $\omega$ and for
      $e=0.5$, 0.8, 0.9, 0.95 (from \textit{left} to \textit{right}).
      We show the errors obtained on the parameters using the approximations $C=1/4$,
      and $C(\hat{\hat{\omega}})$ (see Sect.~\ref{sec:eccentricity}).
      For the approximation $C(\hat{\hat{\omega}})$,
      we also show the errors after one and two iterations of
      the Newton-Raphson correcting algorithm (see Sect.~\ref{sec:refine}).}
    \label{fig:I}
  \end{figure*}
}

\newcommand\figII{
  \begin{figure}
    \centering
    \includegraphics[width=\linewidth]{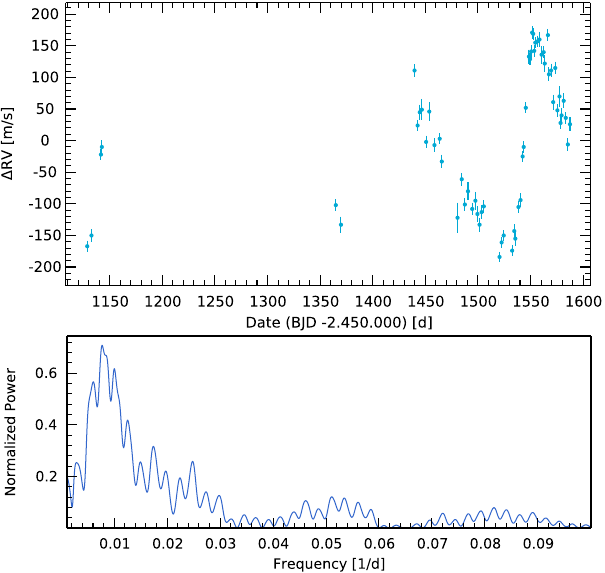}
    \caption{Radial-velocity data of \object{GJ~3021} (\textit{top}) and corresponding
      periodogram (\textit{bottom}).
      The data are taken from \citet{naef_coralie_2001} (61 CORALIE-98 measurements).}
    \label{fig:II}
  \end{figure}
}

\newcommand\figIII{
  \begin{figure}
    \centering
    \includegraphics[width=\linewidth]{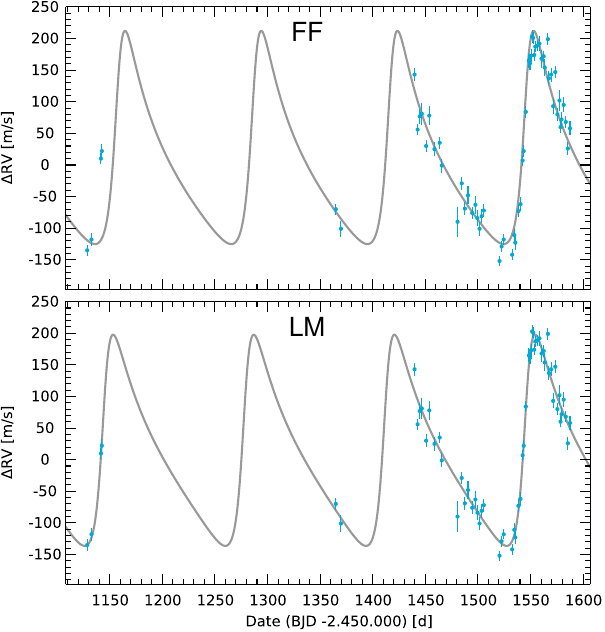}
    \caption{Radial-velocity data of \object{GJ~3021} superimposed with
      the Keplerian model obtained with the FF algorithm (\textit{top}),
      and the model obtained with the LM algorithm starting from the FF solution (\textit{bottom}).
      The orbital elements corresponding to both solutions are presented in Table~\ref{tab:I},
      and compared to the published solution \citep{naef_coralie_2001}.}
    \label{fig:III}
  \end{figure}
}

\newcommand\figIV{
  \begin{figure}
    \centering
    \includegraphics[width=\linewidth]{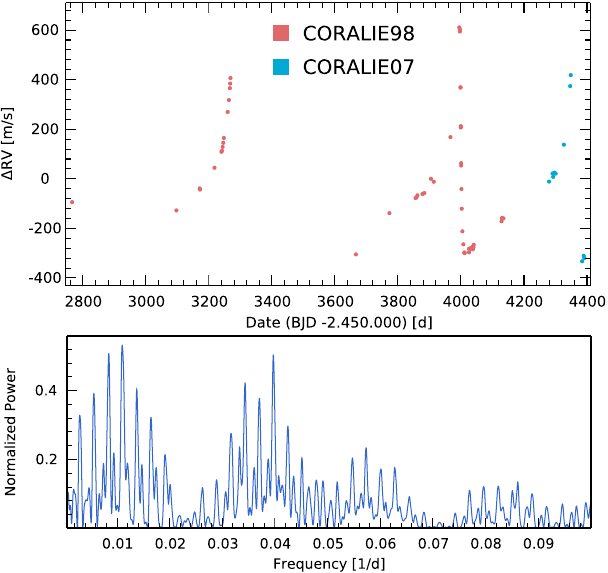}
    \caption{Same as Fig.~\ref{fig:II} but for \object{HD~156846}.
      The data are taken from \citet{tamuz_coralie_2008}.
      They consist of 53 CORALIE-98 measurements (red),
      and  11 CORALIE-07 measurements (blue).}
    \label{fig:IV}
  \end{figure}
}

\newcommand\figV{
  \begin{figure}
    \centering
    \includegraphics[width=\linewidth]{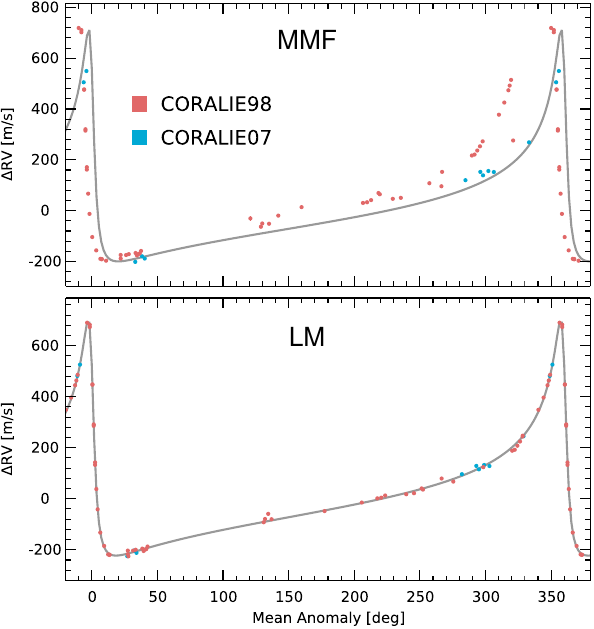}
    \caption{Phase-folded radial-velocity curve of \object{HD~156846}b superimposed with
      the Keplerian model that was obtained with the MMF algorithm (\textit{top}),
      and the model obtained with the LM algorithm starting from the MMF solution (\textit{bottom}).
      The orbital elements corresponding to both solutions are presented in Table~\ref{tab:II},
      and compared to the published solution \citep{tamuz_coralie_2008}.}
    \label{fig:V}
  \end{figure}
}

\newcommand\figVI{
  \begin{figure}
    \centering
    \includegraphics[width=\linewidth]{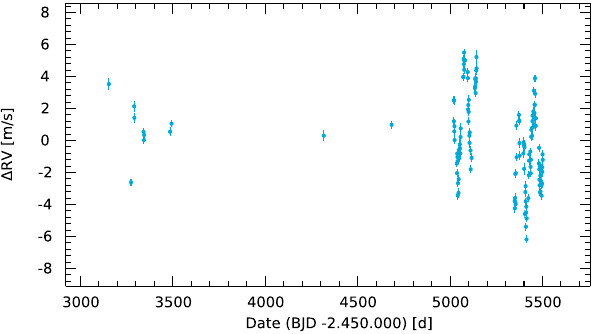}
    \caption{Radial-velocity data of \object{HD~192310}.
      The data are taken from \citet{pepe_harps_2011} (139 HARPS measurements).}
    \label{fig:VI}
  \end{figure}
}

\newcommand\figVII{
  \begin{figure}
    \centering
    \includegraphics[width=\linewidth]{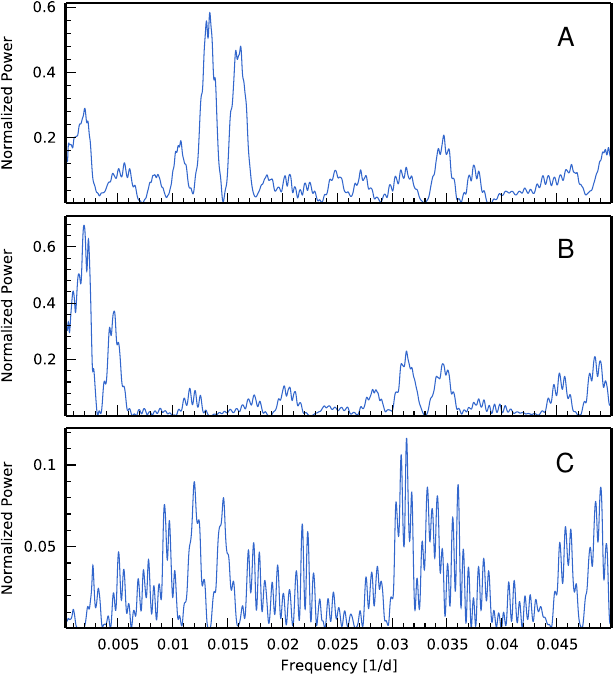}
    \caption{Periodogram of the radial-velocities of \object{HD~192310} (\textbf{A}),
      of the residuals of the one-Keplerian model (LM1 see Table~\ref{tab:III}, \textbf{B}),
      and of the residuals of the two-Keplerians model (LM2 see Table~\ref{tab:III}, \textbf{C}).
      The orbital elements corresponding to the successive solutions are presented in Table~\ref{tab:III},
      and compared to the published solution \citep{pepe_harps_2011}.
    }
    \label{fig:VII}
  \end{figure}
}

\newcommand\figVIII{
  \begin{figure}
    \centering
    \includegraphics[width=\linewidth]{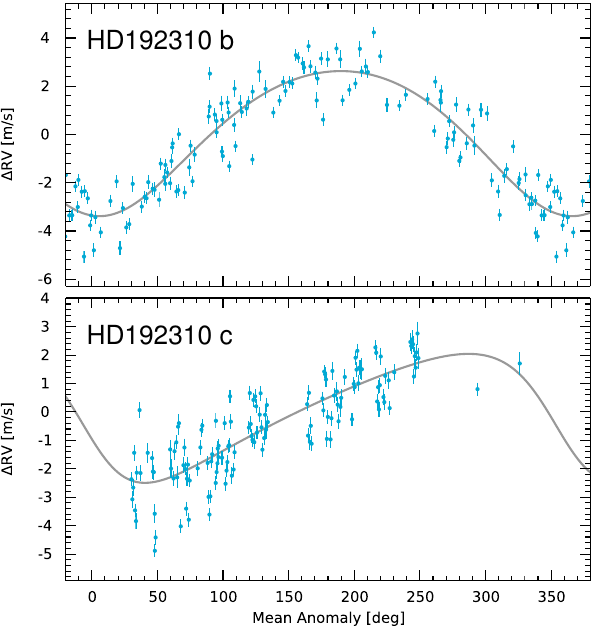}
    \caption{Phase folded radial-velocity curves of \object{HD~192310}b (\textit{top}),
      and c (\textit{bottom}),
      superimposed with our final two-Keplerians model (LM2).
      See Table~\ref{tab:III} for the corresponding orbital elements.}
    \label{fig:VIII}
  \end{figure}
}

\newcommand\figIX{
  \begin{figure}
    \centering
    \includegraphics[width=\linewidth]{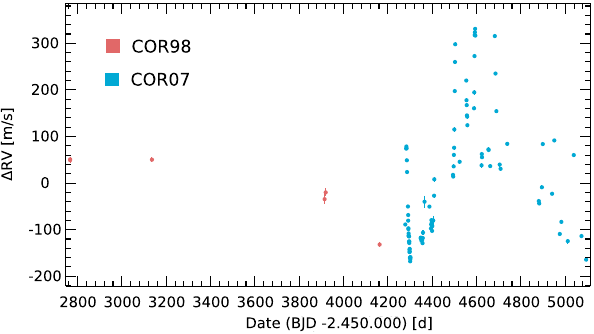}
    \caption{Same as Fig.~\ref{fig:VI} but for \object{HD~147018}.
      The data are taken from \citet{segransan_coralie_2010} and consist of
      6 CORALIE-98 measurements (red) and 95 CORALIE-07 measurements (blue).}
    \label{fig:IX}
  \end{figure}
}

\newcommand\figX{
  \begin{figure}
    \centering
    \includegraphics[width=\linewidth]{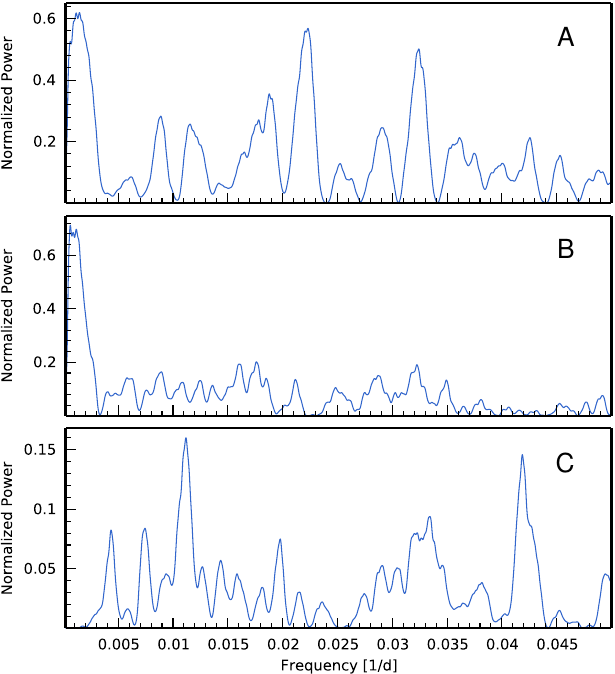}
    \caption{Same as Fig.~\ref{fig:VII} but for \object{HD~147018}.
      The orbital elements corresponding to the successive solutions are presented in Table~\ref{tab:IV},
      and compared to the published solution \citep{segransan_coralie_2010}.}
    \label{fig:X}
  \end{figure}
}

\newcommand\figXI{
  \begin{figure}
    \centering
    \includegraphics[width=\linewidth]{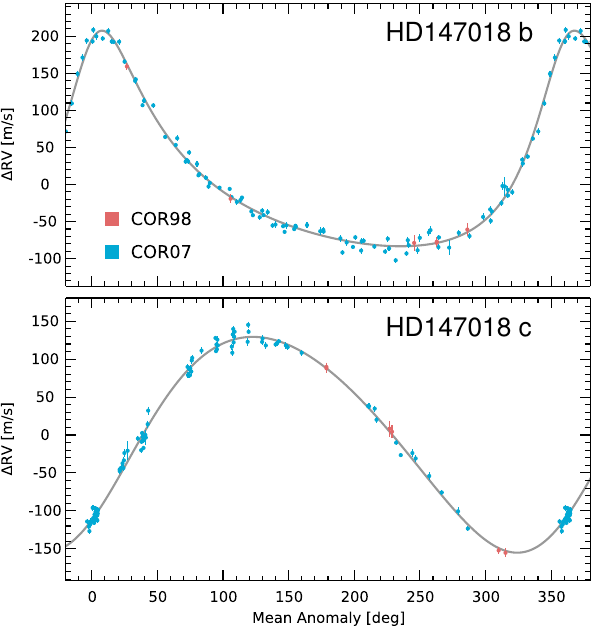}
    \caption{Same as Fig.~\ref{fig:VIII} but for \object{HD~147018}.
      See Table~\ref{tab:IV} for the corresponding orbital elements.}
    \label{fig:XI}
  \end{figure}
}

\newcommand\figAI{
  \begin{figure}
    \centering
    \includegraphics[width=\linewidth]{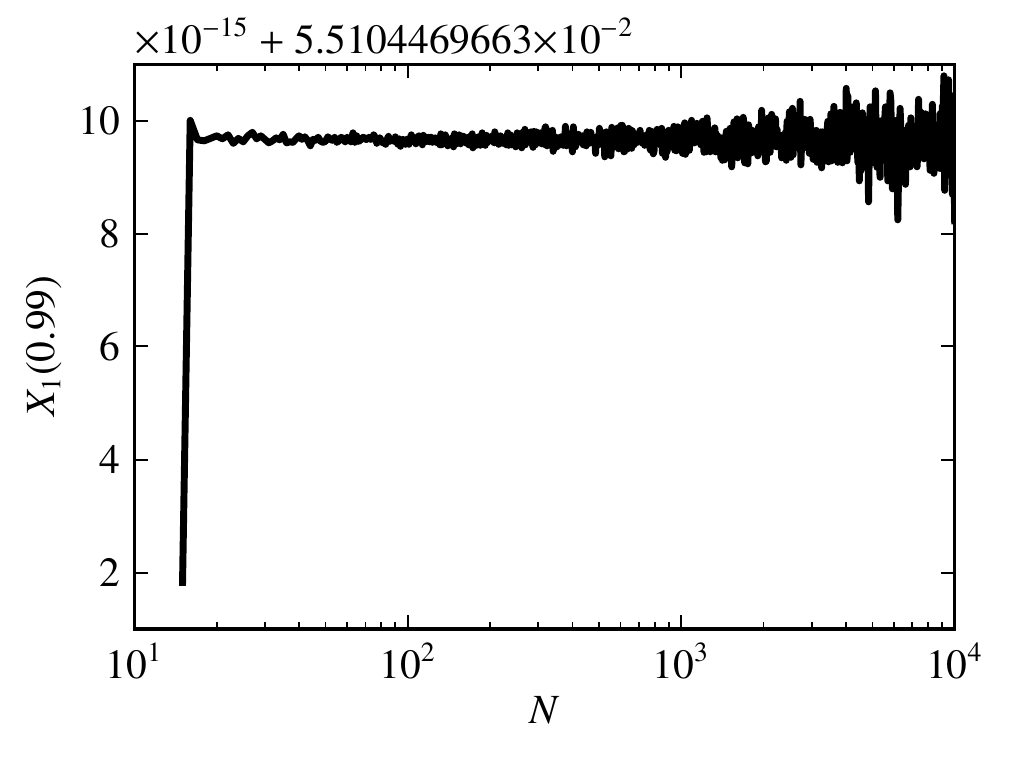}
    \caption{Convergence of the numerical estimation of $X_1(0.99)$.
      We observe that the optimum value of $N$ (see Eq.~(\ref{eq:numHansen})) is about 100.
      For too small $N$ ($\lesssim 20$), the rectangle method does not provide enough precision.
      For too high $N$ ($\gtrsim 300$), the machine errors accumulate and the precision decreases.}
    \label{fig:AI}
  \end{figure}
}

\newcommand\tabI{
  \begin{table*}
    \begin{center}
      \caption{Orbital parameters of \object{GJ~3021}b obtained using our FF algorithm,
        using the LM refinement algorithm (starting from the FF solution),
        and compared with the published solution \citep[PUB,][on the same 61 CORALIE-98 measurements]{naef_coralie_2001}.}
      \begin{tabular}{cc|ccc}
        \hline
        Parameter & [unit] & FF & LM & PUB\\
        \hline
        $\gamma$ & [km/s] &
        $-5.806$ & $-5.806 \pm 0.0027$ & $-5.806 \pm 0.003$\\
        \hline
        $P$ & [day] &
        $129.48$ & $133.72 \pm 0.19$ & $133.71\pm 0.20$ \\
        $K$ & [m/s] &
        $168.7 $ & $167.0 \pm 3.8$ & $167 \pm 4$\\
        $e$ & &
        $0.510$ & $0.514 \pm 0.017$ & $0.511 \pm 0.017$\\
        $\omega$ & [$^\circ$] &
        $-59.6$ & $-69.1 \pm 2.9$ & $-69.3 \pm 3.0$\\
        $T_p$ & [BJD$-2.45\times10^6$] &
        $1546.20$ & $1545.91 \pm 0.63$ & $1545.86 \pm 0.64$\\
        \hline
        $\chi_r^2$ & &
        $9.13$ & $2.50$ & --\\
        \hline
      \end{tabular}
      \label{tab:I}
    \end{center}
  \end{table*}
}

\newcommand\tabII{
  \begin{table*}
    \begin{center}
      \caption{Same as Table~\ref{tab:I} but for \object{HD~156846}b.
        The published solution (PUB) and the data (53 CORALIE-98 measurements, and  11 CORALIE-07 measurements) are taken from \citet{tamuz_coralie_2008}.
        In opposite to \citet{tamuz_coralie_2008},
        we consider here CORALIE-98 and CORALIE-07 as two different instruments,
        and fit for the instrumental offsets individually.}
      \begin{tabular}{cc|ccc}
        \hline
        Parameter & [unit] & MMF & LM & PUB\\
        \hline
        $\gamma_{C98}$ & [km/s] &
        $-68.5659$ & $-68.5372 \pm 0.0015$ & $-68.540 \pm 0.001$\\
        $\gamma_{C07}$ & [km/s] &
        $-68.5544$ & $-68.5307 \pm 0.0023$ & --\\
        \hline
        $P$ & [day] &
        $348.26$ & $359.42 \pm 0.11$ & $359.51 \pm 0.09$ \\
        $K$ & [m/s] &
        $457.8$ & $463.4 \pm 2.5$ & $464.3 \pm 3.0$\\
        $e$ & &
        $0.8523$ & $0.8475 \pm 0.0013$ & $0.8472 \pm 0.0016$\\
        $\omega$ & [$^\circ$] &
        $48.59$ & $52.23 \pm 0.38$ & $52.23 \pm 0.41$\\
        $T_p$ & [BJD$-2.45\times10^6$] &
        $4004.216$ & $3998.078 \pm 0.044$ & $3998.09 \pm 0.05$\\
        \hline
        $\chi_r^2$ & &
        $939$ & $1.28$ & --\\
        \hline
      \end{tabular}
      \label{tab:II}
    \end{center}
  \end{table*}
}

\newcommand\tabIII{
  \begin{table*}
    \begin{center}
      \caption{Orbital parameters of \object{HD~192310}b,~c
        obtained using our FF algorithm,
        using the LM refinement algorithm (starting from the FF solution),
        and compared with the published solution
        \citep[PUB,][on the same 139 HARPS measurements]{pepe_harps_2011}.
        We first fitted the data using our FF algorithm for the inner planet (FF1).
        Then we refined this single-planet solution with the LM method (LM1).
        We added the outer planet using our algorithm on the residuals (FF2),
        and refined the global solution with the LM algorithm (LM2).
        The epoch (used to define the longitudes $\lambda$) is set to 2~455~151.02574596~BJD
        \citep[following][]{pepe_harps_2011}.
      }
      \begin{tabular}{ccc|ccccc}
        \hline
        & Parameter & [unit] & FF1 & LM1 & FF2 & LM2 & PUB\\
        \hline
        & $\gamma$ & [km/s]
        & $-54.22374$ & $-54.22373 \pm 0.00015$ &
        $-54.22343$ & $-54.22319 \pm 0.00027$ & $-54.2232 \pm 0.0003$ \\
        \hline
        & $P$ & [day] &
        $74.66$ & $74.65 \pm 0.11$ & -- & $74.716\pm 0.098$ & $74.72\pm 0.10$ \\
        & $K$ & [m/s] &
        $3.04$ & $3.01 \pm 0.21$ & -- &$3.01 \pm 0.12$ & $3.00 \pm 0.12$\\
        b & $e$ &  &
        $0.214$ & $0.167 \pm 0.069$ & -- &$0.127 \pm 0.039$ & $0.13 \pm 0.04$\\
        & $\omega$ & [$^\circ$] &
        $186$ & $179 \pm 25$ & -- & $172 \pm 21$ & $173 \pm 20$\\
        & $\lambda$ & [$^\circ$] &
        $339.1$ & $338.9 \pm 4.0$ & -- & $340.8 \pm 2.3$ & $340.8 \pm 2.3$\\
        \hline
        & $P$ & [day] &
        -- & -- & $516.6$ & $525.6 \pm 9.5$ & $525.8 \pm 9.2$ \\
        & $K$ & [m/s] &
        -- & -- & $2.07$ & $2.27 \pm 0.27$ & $2.27 \pm 0.28$\\
        c & $e$ &  &
        -- & -- & $0.20$ & $0.31 \pm 0.11$ & $0.32 \pm 0.11$\\
        & $\omega$ & [$^\circ$] &
        -- & -- & $100$ & $109 \pm 21$ & $110 \pm 21$\\
        & $\lambda$ & [$^\circ$] &
        -- & -- & $14$ & $3 \pm 13$ & $1.9 \pm 12.3$\\
        \hline
        & $\chi_r^2$ &  &
        $4.56$ & $4.53$ & $1.43$ & $1.36$ & --\\
        \hline
      \end{tabular}
      \label{tab:III}
    \end{center}
  \end{table*}
}

\newcommand\tabIV{
  \begin{table*}
    \begin{center}
      \caption{Same as Table~\ref{tab:III} but for \object{HD~147018}b,~c.
        The published solution (PUB) is taken from \citet{segransan_coralie_2010}
        as well as the data (6 CORALIE-98 measurements and 95 CORALIE-07 measurements).}
      \begin{tabular}{ccc|ccccc}
        \hline
        & Parameter & [unit] & FF1 & LM1 & FF2 & LM2 & PUB\\
        \hline
        & $\gamma_{C98}$ & [km/s]
        & $-27.391$ & $-27.418 \pm 0.056$ &
        $-27.431$ & $-27.391 \pm 0.014$ & $-27.343 \pm 0.011$ \\
        & $\gamma_{C07}$ & [km/s]
        & $-27.3853$  & $-27.3877 \pm 0.0090$ &
        $-27.3643$ & $-27.3986 \pm 0.0044$ & $-27.348 \pm 0.005$  \\
        \hline
        & $P$ & [day] &
        $44.852$ & $44.756 \pm 0.069$ & -- & $44.2325\pm 0.0079$ & $44.236\pm 0.008$ \\
        & $K$ & [m/s] &
        $167$ & $166 \pm 15$ & -- &$145.5 \pm 1.6$ & $145.33 \pm 1.66$\\
        b & $e$ &  &
        $0.351$ & $0.331 \pm 0.078$ & -- &$0.4674 \pm 0.0076$ & $0.4686 \pm 0.0081$\\
        & $\omega$ & [$^\circ$] &
        $18.5$ & $9 \pm 13$ & -- & $-23.9 \pm 1.3$ & $-24.03 \pm 1.23$\\
        & $T_p$ & [BJD$-2.45\times10^6$] &
        $4461.9$ & $4460.8 \pm 1.4$ & -- & $4459.51 \pm 0.11$ & $4459.49 \pm 0.10$\\
        \hline
        & $P$ & [day] &
        -- & -- & $850$ & $1012 \pm 23$ & $1008 \pm 18$ \\
        & $K$ & [m/s] &
        -- & -- & $95.7$ & $142.3 \pm 4.5$ & $141.2 \pm 4.1$\\
        c & $e$ &  &
        -- & -- & $0.047$ & $0.130 \pm 0.013$ & $0.133 \pm 0.011$\\
        & $\omega$ & [$^\circ$] &
        -- & -- & $-202.9$ & $-134.2 \pm 6.0$ & $-133.1 \pm 6.9$\\
        & $T_p$ & [BJD$-2.45\times10^6$] &
        -- & -- & $4196$ & $4289 \pm 17$ & $4293 \pm 22$\\
        \hline
        & $\chi_r^2$ &  &
        $192$ & $186$ & $58$ & $1.65$ & --\\
        \hline
      \end{tabular}
      \label{tab:IV}
    \end{center}
  \end{table*}
}

\begin{document}

  \title{Analytical determination of orbital elements using Fourier analysis}
  \subtitle{I. The radial velocity case}
  \author{J.-B. Delisle\inst{1,2}
    \and D. S\'egransan \inst{1}
    \and N. Buchschacher \inst{1}
    \and F. Alesina \inst{1}}
  \institute{Observatoire de l'Université de Genève, 51 chemin des Maillettes, 1290, Sauverny, Switzerland\\
    \email{jean-baptiste.delisle@unige.ch}
    \and ASD, IMCCE-CNRS UMR8028, Observatoire de Paris, UPMC,
    77 Av. Denfert-Rochereau, 75014~Paris, France
  }

  \date{\today}

  \abstract{
    We describe an analytical method for computing the orbital parameters
    of a planet from the periodogram of a radial velocity signal.
    The method is very efficient and provides a good approximation of the orbital parameters.
    The accuracy is mainly limited by the accuracy of the computation of the Fourier decomposition of
    the signal which is sensitive to sampling and noise.
    Our method is complementary with more accurate (and more expensive in computer time)
    numerical algorithms (e.g. Levenberg-Marquardt, Markov chain Monte Carlo, genetic algorithms).
    Indeed, the analytical approximation can be used as an initial condition to
    accelerate the convergence of these numerical methods.
    Our method can be applied iteratively to search for multiple planets in the same system.
  }


  \maketitle

  \section{Introduction}
  \label{sec:introduction}

  In this article we are interested in  retrieving planetary orbital parameters
  from a radial velocity time series.
  These parameters are usually obtained using numerical least-squares minimization methods
  (Levenberg-Marquardt, Markov chain Monte Carlo, genetic algorithms, etc.).
  These methods enable one to explore the parameter space
  and find the best fitting solution as well as estimates of the error made in the parameters.
  However, numerical methods need to be initialized with starting values for the parameters.
  When initial values are far from the solution, these algorithms can become
  very expensive in terms of computation time, or even unable to converge
  (non-linear fit can have several local $\chi_2$ minima, etc.).
  On the contrary, using a good guess of the parameters as initial conditions
  significantly improves  the efficiency of numerical methods.

  Here we describe an analytical method that allows a very efficient determination
  of the orbital parameters.
  This method is based on the Fourier decomposition of the radial velocity data, which
  is obtained using linear least-squares spectral analysis.
  This idea of recovering the orbital parameters of the planet from the Fourier decomposition
  of the radial velocity signal has already been proposed by \citet{correia_determination_2008}
  and used in the analysis of different planetary systems
  \citep[see][]{correia_coralie_2005,correia_elodie_2008,correia_harps_2009,correia_harps_2010}.
  A similar idea was also developed in the case of interferometric astrometry
  by \citet{konacki_frequency_2002}.
  In this article, we complete the sketch proposed by \citet{correia_determination_2008}  and provide a fully analytical method to retrieve the orbital parameters.
  In addition to this Fourier method,
  we propose an alternative algorithm that is based on  information contained in the extrema
  of the radial velocity curve and that can be used when the Fourier decomposition is unreliable
  (e.g. ill-sampled, very eccentric planets).
  These algorithms are implemented in the DACE web-platform
  \citep[see][]{buchschacher_DACE_2015}%
  \footnote{The DACE platform is available at \href{http://dace.unige.ch}{http://dace.unige.ch}. The algorithms described in this article are part of the project ``Observations'' of DACE.}.

  In Sect.~\ref{sec:fourier} we derive an analytical decomposition of a Keplerian radial velocity signal
  in Fourier series.
  In Sect.~\ref{sec:F2O}, we describe an analytical method to compute orbital parameters from the Fourier coefficients
  of the fundamental and first harmonics of a Keplerian signal.
  In Sect.~\ref{sec:periodo}, we show how to find the fundamental frequency and compute the Fourier coefficients.
  In Sect.~\ref{sec:examples}, we illustrate the performances and limitations of our methods on observed planetary systems.
  In Sect.~\ref{sec:discussion}, we summarize and discuss our results.

  \section{Fourier decomposition of a Keplerian radial velocity signal}
  \label{sec:fourier}

  We assume here that the observed system is composed of a star and a planet (no perturbation).
  The motion of the star with respect to the center of mass is Keplerian.
  The radial velocity of the star in the reference frame of the center of mass reads
  \begin{equation}
    \label{eq:rv1}
    V(t) = K \left(\cos(\nu+\omega) + e\cos(\omega)\right),
  \end{equation}
  with
  \begin{equation}
    \label{eq:defK}
    K = \frac{m_p}{m_s+m_p}\frac{2\pi a \sin i}{P \sqrt{1-e^2}},
  \end{equation}
  and $a$ is the semi-major axis of the planet with respect to the star,
  $e$ the eccentricity,
  $i$ the inclination,
  $\nu$ the true anomaly,
  $\omega$ the argument of periastron,
  $P$ the orbital period,
  $m_p$ the planet mass,
  $m_s$ the star mass.
  We denote by $n$ the mean-motion of the planet:
  \begin{equation}
    \label{eq:defn}
    n = \frac{2\pi}{P} = \sqrt{\frac{\mu}{a^3}},
  \end{equation}
  with
  \begin{equation}
    \label{eq:defmu}
    \mu = \mathcal{G} (m_s + m_p),
  \end{equation}
  and $\mathcal{G}$ is the gravitational constant.
  The radial velocity signal is $P$-periodic and can be decomposed in discrete Fourier series:
  \begin{equation}
    \label{eq:FourV}
    V(t) = \sum_{k\in \mathbb{Z}} V_k \expo{\i k n t},
  \end{equation}
  with
  \begin{eqnarray}
    V_k &=& \frac{1}{P} \int_0^P V(t) \expo{-\i k n t} \d t\nonumber\\
    \label{eq:Vk}
    &=& \frac{1}{2\pi} \int_0^{2\pi} V(t) \expo{-\i k (M-M_0)} \d M,
  \end{eqnarray}
  where $M$ is the mean anomaly and $M_0$ is the mean anomaly at
  the reference time $t=0$.
  The coefficients $V_k$ are complex numbers.
  Since $V(t)$ is real, we have $V_{-k}=\overline{V}_k$
  (where $\overline{V}_k$ is the complex conjugate of $V_k$).
  We denote by $X_k^{l,q}$ the Hansen coefficients
  \begin{equation}
    \label{eq:hansen}
    X_k^{l,q}(e) = \frac{1}{2\pi} \int_0^{2\pi} \left(\frac{r}{a}\right)^l \expo{\i q \nu} \expo{-\i k M} \d M.
  \end{equation}
  These coefficients are real numbers that only depend on the eccentricity ($e$),
  and we have $X_{-k}^{l,-q}=X_{k}^{l,q}$.
  The Fourier expansion of the radial velocity can be rewritten using Hansen coefficients
  \begin{eqnarray}
    V_0 &=& 0\\
    \label{eq:FourVHansen}
    V_k &=& \frac{K \expo{\i k M_0}}{2}\left(X_k^{0,1}(e)\expo{\i\omega}+X_k^{0,-1}(e)\expo{-\i\omega}\right)\nonumber\\
    &=& \frac{K \expo{\i k M_0}}{2}\left(X_k^{0,1}(e)\expo{\i\omega}+X_{-k}^{0,1}(e)\expo{-\i\omega}\right) \qquad (k \in \mathbb{Z}^*).
  \end{eqnarray}
  We observe that the radial velocity only depends on Hansen coefficients of the form
  $X_k^{0,1}$ ($k \in \mathbb{Z}^*$). In the following, we drop the exponents and use the notation
  $X_k\equiv X_k^{0,1}$.

  We note that $V_0=0$ is only valid if the observer is fixed in the reference
  frame of the center of mass.
  If we account for a constant motion of the observed system with respect to the solar system,
  we have $V_0 \neq 0$ but all other coefficients are unaffected.

  \section{From Fourier coefficients to orbital parameters}
  \label{sec:F2O}

  Let us assume that we are able to determine the period of the planet as well as
  the values of $V_1$ (fundamental) and $V_2$ (first harmonics)
  from an observed radial velocity signal.
  We thus have access to five observables (because $V_k$ are complex numbers).
  Therefore, this information should be sufficient to determine the five parameters:
  $P$, $K$, $e$, $\omega$, $M_0$.

  \subsection{Expanding Hansen coefficients}
  \label{sec:expHansen}

  Hansen coefficients are functions of the eccentricity alone
  and can be expanded in power series of the eccentricity.
  Since we restrict our study to the fundamental and the first harmonics,
  we only need the expansion of $X_{\pm1}$ and $X_{\pm2}$.
  We have (see Appendix~\ref{sec:CompHansenCoeff})
  \begin{eqnarray}
    \label{eq:X1p}
    X_1 &=& 1 - e^2 + \O{e^4}\\
    \label{eq:X2p}
    X_2 &=& e - \frac{5}{4} e^3 + \O{e^5}\\
    \label{eq:X1m}
    X_{-1} &=& - \frac{1}{8} e^2 + \frac{1}{48} e^4 + \O{e^6}\\
    \label{eq:X2m}
    X_{-2} &=& -\frac{1}{12} e^3 + \frac{1}{48}e^5 + \O{e^7}.
  \end{eqnarray}

  \subsection{Eccentricity and phase}
  \label{sec:eccentricity}

  Since we assume that the complex coefficients $V_1$, $V_2$ are known,
  the ratio $\rho \equiv V_2/V_1$ is also known.
  From Eq.~(\ref{eq:FourVHansen}), we have
  \begin{equation}
    \label{eq:rV2V1}
    \rho = \frac{V_2}{V_1} = \expo{\i M_0} \frac{X_2}{X_1}
    \left(\frac{1+X_{-2}/X_2\expo{-2\i\omega}}{1+X_{-1}/X_1\expo{-2\i\omega}}\right).
  \end{equation}
  From Eqs.~(\ref{eq:X1p})-(\ref{eq:X2m}) we obtain
  \begin{eqnarray}
    \label{eq:rapX2X1}
    \frac{X_2}{X_1} &=& e - \frac{e^3}{4} + \O{e^5}\\
    \label{eq:rapXm1X1}
    \frac{X_{-1}}{X_1} &=& -\frac{e^2}{8} + \O{e^4}\\
    \label{eq:rapXm2X2}
    \frac{X_{-2}}{X_2} &=& -\frac{e^2}{12} + \O{e^4}.
  \end{eqnarray}
  We thus have
  \begin{equation}
    \label{eq:diffV1V2X1X2}
    \frac{1+X_{-2}/X_2\expo{-2\i\omega}}{1+X_{-1}/X_1\expo{-2\i\omega}} = 1 + \frac{e^2}{24}\expo{-2\i\omega} + \O{e^4},
  \end{equation}
  and
  \begin{equation}
    \label{eq:rV2V1b}
    \rho = \expo{\i M_0} \left(e - \frac{e^3}{4} + \frac{e^3}{24}\expo{-2\i\omega} + \O{e^5}\right).
  \end{equation}
  We introduce the complex coefficient
  \begin{equation}
    \label{eq:Comega}
    C(\omega) = \frac{1}{4} \left(1 - \frac{\expo{-2\i\omega}}{6}\right),
  \end{equation}
  and use, in the following, the approximation
  \begin{equation}
    \label{eq:rV2V1c}
    \rho \approx \expo{\i M_0} \left(e - C(\omega)e^3 \right),
  \end{equation}
  where order 5 (and more) terms are neglected.
  We first suppose that $C(\omega)$ is known
  and show how to determine the eccentricity and the angle $M_0$.
  Using approximation~(\ref{eq:rV2V1c}), we only have to solve a third order polynomial equation
  to obtain the eccentricity
  \begin{equation}
    \label{eq:rV2V1d}
    |\rho| \approx e - \Re(C) e^3,
  \end{equation}
  where $\Re(C)$ denotes the real part of $C$,
  \begin{equation}
    \label{eq:ReC}
    \Re(C) = \frac{1}{4}\left(1 - \frac{\cos(2\omega)}{6}\right) \quad \in \ \left[\frac{5}{24}, \frac{7}{24}\right].
  \end{equation}
  Since we always have $\Re(C)<1/3$,
  there is at most one solution for $e$ in the interval $[0,1]$, and only
  one, provided that $|\rho|\leq 1-\Re(C)$.
  One can easily verify that this solution is given by
  \begin{equation}
    \label{eq:ecc}
    \hat{e} = \frac{2}{\sqrt{3\Re(C)}} \cos\left(\frac{\pi +
      \arccos\left(\frac{3\sqrt{3\Re(C)}}{2} |\rho|\right)}{3}\right),
  \end{equation}
  where the hat denotes an estimation of the considered parameter (here the eccentricity $e$).
  Using the same approximation (see Eq.~(\ref{eq:rV2V1c})),
  the angle $M_0$ is estimated by
  \begin{equation}
    \label{eq:M0}
    \hat{M}_0 = \arg\left(\frac{\rho}{e-Ce^3}\right).
  \end{equation}
  These estimates (Eq.~(\ref{eq:ecc}) and (\ref{eq:M0})) are based on the assumption that $C(\omega)$ is known.
  As a first (very crude) approximation, we can use $C(\omega) \approx 1/4$, since
  $C$ varies in a small disk in the complex plane around $1/4$ (see Eq.~(\ref{eq:Comega})).
  However a better approximation is given by first computing  an estimate of $\omega$ and using
  Eq.~(\ref{eq:Comega}) to obtain $C$.
  We introduce
  \begin{equation}
    \label{eq:eta}
    \eta = \frac{V_2}{V_1^2} = \frac{2}{K}
    \frac{X_2\expo{\i\omega}+X_{-2}\expo{-\i\omega}}{\left(X_1\expo{\i\omega}+X_{-1}\expo{-\i\omega}\right)^2},
  \end{equation}
  which enables us to eliminate $M_0$.
  We know that $X_{-k}/X_k = \O{e^2}$ (see Eqs.~(\ref{eq:rapXm1X1}),~(\ref{eq:rapXm2X2})),
  thus, at leading order in eccentricity
  \begin{equation}
    \label{eq:eta2}
    \eta \approx \frac{2X_2}{KX_1^2} \expo{-\i\omega},
  \end{equation}
  and the phase of $\eta$ provides a very crude estimate of $\omega$
  \begin{equation}
    \label{eq:crudeEstOmega}
    \hat{\hat{\omega}} = - \arg(\eta),
  \end{equation}
  (where the double hat stands for a very crude estimate)
  which can be used to compute the complex coefficient $C$ (Eq.~(\ref{eq:Comega})),
  and the estimates $\hat{e}$, $\hat{M}_0$.

  \subsection{Other parameters}
  \label{sec:others}

  The remaining parameters ($K$ and $\omega$) are easily obtained now that we have an estimation for
  the eccentricity $\hat{e}$ and the mean anomaly at the reference time ($\hat{M}_0$).
  We only consider the coefficient of the fundamental
  \begin{equation}
    V_1 = \frac{K \expo{\i M_0}}{2} \left(X_1(e)
    \expo{\i\omega}+X_{-1}(e) \expo{-\i\omega}\right),
  \end{equation}
  and rewrite it has a function of $K\cos\omega$, $K\sin\omega$
  \begin{eqnarray}
    \displaystyle V_1 = \frac{\expo{\i M_0}}{2}
    &&\left(K\cos\omega \left(X_1(e)+X_{-1}(e)\right)\right.\nonumber\\
    &&\left. + \i K\sin\omega \left(X_1(e)-X_{-1}(e)\right)\right).
  \end{eqnarray}
  Therefore
  \begin{eqnarray}
    \label{eq:Ct}
    K\cos\omega &=& \frac{2}{X_1(e)+X_{-1}(e)}\Re\left(V_1\expo{-\i M_0}\right)\\
    \label{eq:St}
    K\sin\omega &=& \frac{2}{X_1(e)-X_{-1}(e)}\Im\left(V_1\expo{-\i M_0}\right).
  \end{eqnarray}
  We obtain estimates for $K$ and $\omega$ by introducing the estimates of $e$ and $M_0$
  in Eqs.~(\ref{eq:Ct}),~(\ref{eq:St}):
  \begin{eqnarray}
    \label{eq:C}
    \displaystyle \hat{K}\cos\hat{\omega} &=& \frac{2}{X_1(\hat{e})+X_{-1}(\hat{e})}\Re\left(V_1\expo{-\i \hat{M}_0}\right)\\
    \label{eq:S}
    \displaystyle \hat{K}\sin\hat{\omega} &=& \frac{2}{X_1(\hat{e})-X_{-1}(\hat{e})}\Im\left(V_1\expo{-\i \hat{M}_0}\right).
  \end{eqnarray}
  The Hansen coefficients $X_{\pm 1}$ can be evaluated at $\hat{e}$ using the power series expansions of Eqs.~(\ref{eq:X1p}),~(\ref{eq:X1m}) or by using numerical estimations of the integral of Eq.~(\ref{eq:hansen}) (see Appendix~\ref{sec:CompHansenCoeff}).
  In the case of vanishing eccentricity, $M_0$ and $\omega$ are meaningless quantities,
  while the mean longitude at reference time $\lambda_0 = M_0 + \omega$
  is the only important angle.
  Our analytical method provides an accurate estimate for $\lambda_0$
  even in the low eccentricity case.
  Indeed, at low eccentricity,
  the estimate $\hat{M}_0$ (see Eq.~(\ref{eq:M0})) may be incorrect because
  the coefficient $V_2$ is small and more sensitive to noise.
  However, for small $e$, we have $X_{1}(e)\approx 1$ and $X_{-1}(e) \ll X_1(e)$,
  and thus
  \begin{equation}
    V_1 \approx \frac{K}{2} \expo{\i \lambda_0}.
  \end{equation}
  The phase of $V_1$ directly provides an estimate of $\lambda_0$.
  Still assuming small eccentricity, we have (see Eqs.~(\ref{eq:C}), (\ref{eq:S}))
  \begin{equation}
    \hat{K}\expo{\i \hat{\omega}} \approx 2 V_1 \expo{-\i\hat{M}_0}.
  \end{equation}
  Therefore, $\hat{\omega}$ is also incorrect,
  but the estimate $\hat{\lambda}_0 = \hat{M}_0 + \hat{\omega}$ is not affected:
  \begin{equation}
    \hat{K}\expo{\i \hat{\lambda}_0} \approx K\expo{\i \lambda_0}.
  \end{equation}

  In principle, higher order harmonics could be used to check and/or improve the solution.
  In particular, the second harmonics could be used to discriminate
  between two planets in a 2:1 resonance and a single eccentric planet.
  Indeed, if the second harmonics coefficient is not compatible with the orbital
  parameters that were determined from the fundamental and first harmonics,
  it could be  evidence of the presence of a second planet that is in resonance.
  However, the amplitude of the $k$-th harmonics is of order $K e^k$.
  Thus the Fourier coefficients of higher order harmonics
  are much more sensitive     to the noise and less reliable,
  especially for low eccentricities.

  \subsection{Refinement of estimates}
  \label{sec:refine}

  Our analytical determination of the parameters gives very good results for small to moderate eccentricities
  (see Sect.~\ref{sec:accuracy}).
  However, at high eccentricities, the estimates exhibit significant divergences
  from the true parameters (while still being in their neighbourhood).
  To refine the estimates, one can use a simple Newton-Raphson method.
  We denote by $x$ the vector of parameters and $y$ the vector of Fourier coefficients:
  \begin{equation}
    \label{eq:x}
    x =
    \begin{pmatrix}
      K\\
      e\\
      \omega\\
      M_0
    \end{pmatrix},
    \quad
    y =
    \begin{pmatrix}
      \Re(V_1)\\
      \Im(V_1)\\
      \Re(V_2)\\
      \Im(V_2)
    \end{pmatrix}.
  \end{equation}
  We denote by $J$ the Jacobian matrix of our system
  \begin{equation}
    \label{eq:jacob}
    J =
    \begin{pmatrix}
      \displaystyle\frac{\partial y_1}{\partial x_1} & \cdots & \displaystyle\frac{\partial y_1}{\partial x_4}\\
      \vdots & \ddots & \vdots\\
      \displaystyle\frac{\partial y_4}{\partial x_1} & \cdots & \displaystyle\frac{\partial y_4}{\partial x_4}
    \end{pmatrix}.
  \end{equation}
  The Newton-Raphson method provides corrections for the estimates of the parameters $x$
  \begin{equation}
    \label{eq:newtonraphson}
    \delta \hat{x} = J^{-1} (y-\hat{y}).
  \end{equation}
  From our estimates of the parameters, we can compute estimates for $V_1$, $V_2$ (see Eq.~(\ref{eq:FourVHansen})),
  and thus determine the $\hat{y}$ vector and compare it to the known Fourier coefficients (vector $y$).
  The estimation of $V_1$, $V_2$ requires estimates for the Hansen coefficients $X_{\pm1}$, $X_{\pm2}$
  which can be obtained analytically or numerically (see Appendix~\ref{sec:CompHansenCoeff}).
  The computation of the matrix $J$ requires
  estimates of the partial derivatives of $V_1$, $V_2$.
  From Eq.~(\ref{eq:FourVHansen}) we deduce
  \begin{eqnarray}
    \label{eq:dVdK}
    \frac{\partial V_k}{\partial K} &=& \frac{V_k}{K} = \frac{\expo{\i k M_0}}{2}\left(X_k\expo{\i\omega}+X_{-k}\expo{-\i\omega}\right),\\
    \frac{\partial V_k}{\partial e} &=& \frac{K\expo{\i k M_0}}{2}\left(\frac{\d X_k}{\d e}\expo{\i\omega}+\frac{\d X_{-k}}{\d e}\expo{-\i\omega}\right),\\
    \frac{\partial V_k}{\partial \omega} &=& \i \frac{K \expo{\i k M_0}}{2}\left(X_k\expo{\i\omega}-X_{-k}\expo{-\i\omega}\right),\\
    \frac{\partial V_k}{\partial M_0} &=& \i k V_k= \i k \frac{K \expo{\i k M_0}}{2}\left(X_k\expo{\i\omega}+X_{-k}\expo{-\i\omega}\right).
  \end{eqnarray}
  We thus need to determine estimates of the derivatives of Hansen coefficients
  $\displaystyle\frac{\d X_k}{\d e}$ (for $k=-2,-1,1,2$).
  We use the same method as for Hansen coefficients computation (see Appendix~\ref{sec:DerivHansenCoeff}).
  Then, we only need to invert the $4\times 4$ matrix $J$ to obtain the corrections for the orbital parameters.
  This process can be reiterated to increase the accuracy.
  This numerical method converges very rapidly since we already have good approximations for the parameters.

  \subsection{Accuracy of estimates}
  \label{sec:accuracy}

  To test the accuracy of our algorithm,
  we computed the Fourier coefficients $V_1$ and $V_2$
  according to Eq.~(\ref{eq:FourVHansen}) for different values of
  $e$ and $\omega$ and using a numerical computation of the Hansen coefficients (see Appendix~\ref{sec:CompHansenCoeff}).
  We arbitrarily set $M_0 = 0$, and $K=1$ because the value of these parameters should not affect the accuracy
  of the algorithm.
  We then applied our algorithm to obtain estimates of $K$, $e$, $\omega$, $M_0$ and compared them
  to the expected values.
  Figure~\ref{fig:I} shows the results of this comparison, using the estimate $C \approx 1/4$,
  or using $C(\hat{\hat{\omega}})$ (see Sect.~\ref{sec:eccentricity}).
  In the latter case, we also show the results of the Newton-Raphson refinement after one and two iterations.
  We observe that for $e=0.5$, all methods approximate  the orbital parameters very well
  (see Fig.~\ref{fig:I}, \textit{left}).
  Above $0.8$, the estimate using $C\approx 1/4$ does not constrain the eccentricity and the
  argument of periastron well, while the use of $C(\hat{\hat{\omega}})$
  improves the results significantly (see Fig.~\ref{fig:I}).
  Finally, we see that, with only two iterations of the Newton-Raphson correction, we obtain very accurate
  parameters, even for $e=0.95$ (see Fig.~\ref{fig:I}, \textit{right}).
  Therefore, for applications to real data (see Sect.~\ref{sec:examples})
  we systematically use these two steps of Newton-Raphson correction in our algorithm.
  These encouraging results are obtained assuming that the Fourier coefficients $V_1$, $V_2$
  have been determined very accurately.
  In real cases the errors might be much greater if the estimates of
  $V_1$, $V_2$ are not precise enough.

  \figI

  \section{Period finding and Fourier coefficients}
  \label{sec:periodo}

  Our estimation of the planetary orbital parameters is based on the assumption that
  we are able to determine the planet period and the Fourier coefficients $V_1$, $V_2$ (see Eq.~(\ref{eq:Vk}))
  from the radial velocity signal.
  The classical method for this is to use a least-square spectral analysis
  \citep[such as the Lomb-Scargle periodogram, e.g.][]{zechmeister_generalised_2009}.
  We denote by $t_i$, $v_i$, and $\sigma_i$ ($i\in [1,N_\mathrm{m}]$)
  the time, the value of radial velocity, and the estimated (Gaussian) noise of each measurement.
  We assume that
  \begin{equation}
    v_i = V(t_i) + \epsilon_i,
  \end{equation}
  with $\epsilon_i$ randomly drawn from a normal distribution $\mathcal{N}(0,\sigma_i)$.
  We aimed to decompose the radial velocity signal on a basis of functions $f_k(t)~(k\in [1,N_f])$,
  such that
  \begin{equation}
    \label{eq:Decompf}
    V(t) = \sum_{k=1}^{N_f} C_k f_k(t).
  \end{equation}
  To find the coefficients $C_k$, we perform a weighted least-squares regression.
  We introduce the matrices
  \begin{equation}
    v = \begin{pmatrix}
      v_1/\sigma_1\\
      v_2/\sigma_2\\
      \vdots\\
      v_{N_\mathrm{m}}/\sigma_{N_\mathrm{m}}
    \end{pmatrix},
    \qquad  \qquad
    C = \begin{pmatrix}
      C_1\\
      C_2\\
      \vdots\\
      C_{N_f}
    \end{pmatrix},
  \end{equation}
  \begin{equation}
    f = \begin{pmatrix}
      f_1(t_1)/\sigma_1 & f_2(t_1)/\sigma_1 & \cdots & f_{N_f}(t_1)/\sigma_1\\
      f_1(t_2)/\sigma_2 & f_2(t_2)/\sigma_2 & \cdots & f_{N_f}(t_2)/\sigma_2\\
      \vdots  & \vdots  & \ddots & \vdots\\
      f_1(t_{N_\mathrm{m}})/\sigma_{N_\mathrm{m}} & f_2(t_{N_\mathrm{m}})/\sigma_{N_\mathrm{m}}
      & \cdots & f_{N_f}(t_{N_\mathrm{m}})/\sigma_{N_\mathrm{m}}
    \end{pmatrix}.
  \end{equation}
  The least-squares regression provides an estimator for $C$
  \begin{equation}
    \label{eq:invLS}
    \hat{C} = \left(f^{\rm T} f\right)^{-1} f^{\rm T} v.
  \end{equation}

  The functions $f_k$ that are used in the decomposition basis are usually
  the sine and cosine of a given period:
  \begin{eqnarray}
    f_1(t) &=& \cos\left(2\pi \frac{t}{P}\right)\\
    f_2(t) &=& \sin\left(2\pi \frac{t}{P}\right).
  \end{eqnarray}
  In addition to these sinusoidal functions, one can introduce shifts for each instrument (which also account for
  the proper motion of the system with respect to the solar system), as well as
  power law drifts ($t^k$, $k=1,2,3$...) to better model the signal.

  We then compute the squared difference between the model and the data ($\chi^2(P)$ which
  depends on the chosen period $P$).
  The best-fitting period is found by minimizing the $\chi^2(P)$ function.
  One can construct the periodogram of the signal by computing the power $\mathcal{P}(P)$
  for a range of periods \citep[e.g.][]{zechmeister_generalised_2009}.
  The power $\mathcal{P}$ is simply a renormalization of the $\chi^2$ function
  \citep[see][]{zechmeister_generalised_2009}:
  \begin{equation}
    \label{eq:powerP}
    \mathcal{P}(P) = \frac{\chi^2_0 - \chi^2(P)}{\chi^2_0},
  \end{equation}
  where $\chi^2_0$ is the squared difference between the data and the model without the periodic functions (only shifts and drifts).

  We assume that the period of the planet corresponds to a peak in the periodogram
  (which is equivalent to a minimum of $\chi^2$).
  We denote by $\hat{P}$ the period that corresponds with this peak.
  We now need estimates for $V_1$ and $V_2$.
  We obtain these by using the same least-squares algorithm, but we introduce in the function
  basis $f$ the sines and cosines corresponding to both  the fundamental period $\hat{P}$ and the first harmonics $\hat{P}/2$:
  \begin{eqnarray}
    f_1(t) &=& \cos\left(2\pi \frac{t}{\hat{P}}\right)\\
    f_2(t) &=& \sin\left(2\pi \frac{t}{\hat{P}}\right)\\
    f_3(t) &=& \cos\left(4\pi \frac{t}{\hat{P}}\right)\\
    f_4(t) &=& \sin\left(4\pi \frac{t}{\hat{P}}\right).
  \end{eqnarray}
  Again, we can complete this basis with instrumental shifts and drifts and we obtain
  estimates for the coefficients $C_k$ (see Eq.~(\ref{eq:invLS})).
  By definition, $V_1$ and $V_2$ are then given by
  \begin{eqnarray}
    \label{eq:estV1}
    \hat{V}_1 &=& \frac{\hat{C}_1 - \i \hat{C}_2}{2}\\
    \label{eq:estV2}
    \hat{V}_2 &=&  \frac{\hat{C}_3 - \i \hat{C}_4}{2},
  \end{eqnarray}
  and the method described in Sect.~\ref{sec:F2O} can be used
  to retrieve the orbital parameters of the planet from these coefficients.

  As shown in Sect.~\ref{sec:accuracy} (Fig.~\ref{fig:I}), our estimates of orbital parameters
  are very accurate if the Fourier coefficients are perfectly determined.
  In real cases, the accuracy of Fourier coefficients is affected
  by the noise and the sampling of the data.
  The determination of Fourier coefficients is thus the main source of error in our algorithm.
  In particular, for very eccentric planets,
  if the first harmonics amplitude is overestimated (compared to the fundamental amplitude),
  there may not be any meaningful solution for the orbital parameters (especially the eccentricity),
  and our algorithm cannot be used.
  In this case, we can use an alternative method to find estimates for the orbital elements.
  In Appendix~\ref{sec:extrema}, we describe a method that makes use of
  the values and timings of the minimum and maximum of the (folded) radial velocity curve
  to provide rough estimates of the orbital parameters (to use when the Fourier method breaks).

  \section{Illustrations and performances}
  \label{sec:examples}

  In this section, we apply our method to actual radial velocity data
  to illustrate its accuracy and limitations.
  All the algorithms described in this article have been implemented in
  the DACE platform \citep[\href{http://dace.unige.ch}{http://dace.unige.ch}, see][]{buchschacher_DACE_2015}.
  The DACE platform also implements a Levenberg-Marquardt algorithm that enables us to refine
  the solution initially found with our methods.
  We denote by Fourier fit (FF) (see Sect.~\ref{sec:F2O}), min/max fit (MMF) (see Appendix~\ref{sec:extrema}),
  and Levenberg-Marquardt (LM) the different algorithms we use to fit the data and find the orbital elements.
  All the following results are obtained using the DACE platform.
  We selected four representative cases:
  \object{GJ~3021} (one eccentric planet),
  \object{HD~156846} (one ill-sampled, very eccentric planet),
  \object{HD~192310} (two planets with a low SNR),
  and \object{HD~147018} (two planets with equal semi-amplitudes).
  For all these analyses, we quadratically added   an instrumental systematic error to the radial velocity uncertainty.
  For CORALIE-98, this instrumental error is set to 6~m/s,
  for CORALIE-07 we use 5~m/s,
  and for HARPS 0.75~m/s.

  We note that our FF method provides values for $K$, $e$, $\omega$, $M_0$, but
  other sets of parameters can then be derived and used for the LM fit.
  In particular, for low eccentricities and/or low S/N,
  the set $K$, $k=e\cos\omega$, $h=e\sin\omega$, $\lambda_0 = M_0+\omega$
  is better suited.
  Indeed, for low eccentricities, $M_0$ and $\omega$ present strong anti-correlations
  since the periastron direction is ill-defined.
  Using $k$, $h$, $\lambda_0$ allows to avoid this issue.
  In the following illustrations,
  we used the same set of parameters as the published ones to be able to compare our results with published solutions.

  \subsection{GJ~3021}
  \label{sec:gj-3021}

  \figII
  \tabI
  \figIII

  \object{GJ~3021}b is an eccentric ($e \approx 0.5$) planet
  discovered by the CORALIE survey \citep[see][]{naef_coralie_2001}.
  We applied our method (using DACE) to the exact same data as \citet{naef_coralie_2001},
  which consist of 61 CORALIE-98 measurements.
  Figure~\ref{fig:II} shows these data and the corresponding periodogram.
  From this periodogram we selected the period that corresponds to the highest peak
  (about 130 days) and applied our FF algorithm.
  We then used this FF solution as an initial step for an LM algorithm
   to refine our solution.
  The orbital parameters provided by both algorithms (FF and LM) are shown in Table~\ref{tab:I},
  and the corresponding radial velocity curves are shown (superimposed with the data) in Fig.~\ref{fig:III}.

  The LM solution is almost identical to the published solution and the differences are well below the
  error bars (see Table~\ref{tab:I}).
  The FF solution exhibits more significant differences but remains a very good approximation
  (see Table~\ref{tab:I}),
  especially considering the fact that only one period is well sampled (see Fig.~\ref{fig:III}).

  The case of \object{GJ~3021} shows that the FF solution can provide a very good approximation
  of the orbital parameters even for quite high eccentricities ($e\approx 0.5$).
  This approximation can then be refined with numerical methods (such as the LM algorithm),
  which converge much more rapidly to the solution than without this first approximation.

  \subsection{HD~156846}
  \label{sec:hd-156846}

  \figIV
  \tabII
  \figV

  \object{HD~156846}b is a very eccentric planet ($e\approx 0.85$) with
  a period very close to one year ($P\approx 360$~d)
  discovered by the CORALIE survey
  \citep[see][]{tamuz_coralie_2008}.
  We used the same data as \citet{tamuz_coralie_2008} (53 CORALIE-98 measurements, and  11 CORALIE-07 measurements).
  These data are shown in Fig.~\ref{fig:IV}, together with the corresponding periodogram.
  We note that in \citet{tamuz_coralie_2008}, CORALIE-98 and CORALIE-07 were treated as a single instrument because the offset between them is relatively small.
  In our study we release this constraint and consider both instruments individually (for the offsets).

  The proximity of the period to one year induces sampling problems (i.e. a gap in the phase-folded curve).
  Together with the high eccentricity, the sampling issues prevent an accurate determination of
  the Fourier coefficients.
  In particular, the amplitude of the fundamental is underestimated
  (compared to the harmonics, see the periodogram in Fig.~\ref{fig:IV}~\textit{bottom}).
  The highest peak in the periodogram
  (see Fig.~\ref{fig:IV}~\textit{bottom}) corresponds to the third harmonics of the planet
  (period of $P/4$).
  The determination of the correct fundamental period is thus more difficult than in the case
  of \object{GJ~3021} (see Sect.~\ref{sec:gj-3021}).
  However, the peak at the fundamental period is still visible, as well as the peaks
  that correspond to the first and second harmonics.
  One can thus observe that the four peaks are related ($P$, $P/2$, $P/3$, $P/4$), and
  select the correct fundamental period.
  In addition to this issue, even if the correct period is chosen, the
  determined Fourier coefficients are not compatible with a Keplerian signal
  (the ratio between the first harmonics and the fundamental amplitudes is too large).
  The FF method cannot be used in this case.
  As an alternative to the FF method, we used the MMF method, which is
  based on the extrema information (see Appendix~\ref{sec:extrema}).
  This method provides crude estimates of the orbital parameters that we use as an initial
  step for the LM algorithm.
  Table~\ref{tab:II} shows the orbital elements provided by the MMF and LM algorithms compared with
  the published ones \citep[from][]{tamuz_coralie_2008}.
  Figure~\ref{fig:V} shows the phase-folded data superimposed with both solutions (MMF and LM).
  In Table~\ref{tab:II} and Fig.~\ref{fig:V}, we observe that the orbital parameters provided by the
  MMF algorithm are a good first approximation.
  The main issue is in the determination of instrumental offsets
  (see Fig.~\ref{fig:V}).
  Finally, the LM solution is consistent with the published solution (see Table~\ref{tab:II}).

  The case of \object{HD~156846} shows that sampling issues (associated with high eccentricity)
  may prevent the use of the FF algorithm.
  In such cases, the MMF solution seems to provide acceptable estimates for the orbital parameters.
  These estimates provide a very good initialization step for the LM algorithm.

  \subsection{HD~192310}
  \label{sec:hd-192310}

  \figVI
  \figVII
  \tabIII
  \figVIII

  \object{HD~192310} hosts two planets discovered with HARPS \citep[see][]{pepe_harps_2011}.
  We use the 139 HARPS measurements provided by \citet{pepe_harps_2011} for our analysis.
  These data are shown in Fig.~\ref{fig:VI} and the corresponding periodogram is shown in
  Fig.~\ref{fig:VII} (panel \textbf{A}).
  The semi-amplitudes of the radial velocity signal corresponding to both planets are small
  \citep[about 3~m/s, see][]{pepe_harps_2011}.
  Thus, the radial velocity signal has a low S/N, which is challenging for our algorithm.
  We applied the FF method on the inner planet (b) first,
  which provided us with a first solution FF1.
  We then refined this solution with the LM algorithm (LM1).
  Figure~\ref{fig:VII} (panel \textbf{B}) shows the periodogram of the residuals of this one-planet solution (LM1).
  We then applied  the FF algorithm again on the residuals to obtain the orbital parameters of  Planet c (FF2).
  Finally, we refined the global solution (Planets b and c) using the LM method (LM2).
  Figure~\ref{fig:VII} (panel \textbf{C}) shows the periodogram of the residuals of this two-planet solution (LM2).
  The four successive solutions (FF1, LM1, FF2, and LM2) are shown in Table~\ref{tab:III}
  and compared with the published solution \citep[PUB,][]{pepe_harps_2011}.
  The phase-folded radial velocity data superimposed with the final model (LM2) are shown in Fig.~\ref{fig:VIII}.

  The final LM2 solution agrees very well with the published solution (see Table~\ref{tab:III}).
  The intermediate solutions (FF1, LM1, and FF2) exhibit larger differences but remain
  close to the LM2 solution.
  The most significant differences is observed in the eccentricities.
  The eccentricity of Planet b is slightly overestimated in FF1 and LM1,
  while the eccentricity of Planet c is slightly underestimated in FF2 (see Table~\ref{tab:III}).
  The complete fit (LM2) is necessary to solve these issue and refine the final model.

  The case of \object{HD~192310} shows that the FF method can be used even for
  multi-planetary systems with low S/N, and still provides good estimates for the orbital parameters.

  \subsection{HD~147018}
  \label{sec:hd-147018}

  \figIX
  \figX
  \tabIV
  \figXI

  \object{HD~147018} hosts two planets discovered by the CORALIE survey
  \citep[see][]{segransan_coralie_2010}.
  We analyzed the same data as \citet{segransan_coralie_2010} (6 CORALIE-98 and 95 CORALIE-07 measurements), which are shown in Fig.~\ref{fig:IX}.
  The periodogram is shown in Fig.~\ref{fig:X} (panel \textbf{A}).
  Both planets have equivalent semi-amplitudes \citep[$\approx 140$ m/s, see][]{segransan_coralie_2010}.
  We applied the same steps as for \object{HD~192310} (see Sect.~\ref{sec:hd-192310}).
  The four successive solutions (FF1, LM1, FF2, and LM2) are shown in Table~\ref{tab:IV}.
  The periodogram of the residuals of solution LM1 and LM2 are shown in Fig.~\ref{fig:X} (panels \textbf{B} and \textbf{C}).
  The phase-folded radial velocity data superimposed with the LM2 model is shown in Fig.~\ref{fig:XI}.

  The final LM2 solution is consistent with the published solution (see Table~\ref{tab:IV}).
  The intermediate solutions (FF1, LM1, and FF2) exhibit significant differences
  (e.g. on the outer period or the angles $\omega$, see Table~\ref{tab:IV}).
  This is probably due to the iterative method we used:
  the planets are fitted one after the other by looking at the residuals of the previous model.
  The case of \object{HD~147018} is representative of the worst situation for such an iterative method.
  Indeed, both planets have very similar semi-amplitudes so they both contribute to the same amount in the radial velocity signal.
  When fitting the data for the first planet (FF1, LM1),
  the second planet signal acts as a noise (which, in the case of \object{HD~147018}, has the amplitude as the first planet signal).

  The case of \object{HD~147018} shows that the FF method alone cannot be used to reliably
  fit multi-planetary systems with equivalent semi-amplitudes.
  However it is still a good first approximation to use as an initial step
  for an overall (all planets at a time) numerical fit (e.g. LM).

  \section{Discussion}
  \label{sec:discussion}

  In this article we describe an analytical method to retrieve the orbital parameters
  of a planet from the periodogram of a radial velocity signal.
  The method can be outlined as follows:
  \begin{enumerate}
    \item Find the period ($P$) of the planet from the periodogram
    \item Compute the Fourier coefficients of the fundamental and the first harmonics ($V_1$, $V_2$, see Eqs.~(\ref{eq:estV1}),~(\ref{eq:estV2}))
    \item Compute the orbital parameters ($e$, $M_0$, $K$, $\omega$) from these Fourier coefficients (see Eqs.~(\ref{eq:ecc}),~(\ref{eq:M0}),~(\ref{eq:C}),~(\ref{eq:S}))
    \item Possibly refine the parameters with a numerical method
    \item Compute the residuals and reiterate to find other planets
  \end{enumerate}

  This analytical method is very efficient for retrieving  parameters compared to numerical methods.
  The determination of the orbital parameters from the Fourier coefficients $V_1$, $V_2$ is very
  accurate, even at high eccentricities (see Fig.~\ref{fig:I}).
  The main limitation of the technique comes from the estimation of the Fourier coefficients.
  The accuracy of these coefficients (and thus of the final parameters)
   depends greatly on the sampling of the signal.
  In some ill-sampled cases, the inaccuracy of the Fourier coefficients prevents the use of our method (see Sect.~\ref{sec:hd-156846}).
  In such cases, an alternative method that is based on the extrema information can be used to obtain
  estimates for the orbital parameters (see Appendix~\ref{sec:extrema}).

  Our analytical method is complementary with numerical algorithms since it provides
  very good initialization values for them.
  We have illustrated this complementarity by using the Levenberg-Marquardt numerical
  method in combination with our analytical algorithm (see Sect.~\ref{sec:examples}).
  In particular, for multi-planetary cases,
  a global numerical fit can significantly improve the solution,
  especially when the semi-amplitudes of the different planets are comparable
  (see Sect.~\ref{sec:hd-147018}).
  We emphasize the fact that the aim of this method is not to replace more complex numerical algorithms,
  but to complement them.
  Our method can be used in combination
  with any numerical method (Levenberg-Marquardt, Markov chain Monte Carlo, genetic algorithms, etc.),
  and can significantly reduce the parameter space that needs to be explored,
  and improve the convergence speed.
  This is especially relevant for multi-planetary systems for which the parameter space
  has many dimensions and the convergence of numerical algorithm can be slow.

  Our methods are all implemented in the DACE platform \citep{buchschacher_DACE_2015},
  available at \href{https://dace.unige.ch}{http://dace.unige.ch} (under the project Observations).

  \begin{acknowledgements}
    We thank the anonymous referee for his/her useful comments.
    This work has, in part, been carried out within the framework of
    the National Centre for Competence in Research PlanetS
    supported by the Swiss National Science Foundation.
    The authors acknowledge financial support from the SNSF.
    This publication makes use of
    DACE, a Data Analysis Center for Exoplanets,
    a platform of the Swiss National Centre of Competence in Research (NCCR) PlanetS,
    based at the University of Geneva (CH).
  \end{acknowledgements}

  \appendix

  \section{Computation of Hansen coefficients}
  \label{sec:CompHansenCoeff}

  In this appendix, we show how to compute Hansen coefficients, both analytically (power series of eccentricity)
  and numerically.
  There is a very broad literature on this topic \citep[e.g.][]{laskar_note_2005,laskar_explicit_2010}.
  Here we simply aim at illustrating how to compute
  a sub-family of Hansen coefficients: $X_k^{0,1}$ ($k\in\mathbb{Z}$).
  In this restricted case, the definition of Eq.~(\ref{eq:hansen}) rewrites
  \begin{equation}
    \label{eq:restrichansen}
    X_k = X_k^{0,1} = \frac{1}{2\pi} \int_0^{2\pi} \expo{\i \nu} \expo{-\i k M} \d M.
  \end{equation}
  Using
  \begin{eqnarray}
    M &=& E - e\sin E,\\
    \d M &=& \frac{r}{a} \d E,\\
    \expo{\i\nu} &=& \frac{a}{r} \left(\cos E-e + \i \sqrt{1-e^2} \sin E\right),
  \end{eqnarray}
  we obtain
  \begin{equation}
    \label{eq:hansenE}
    X_k = \frac{1}{2\pi} \int_0^{2\pi}  \left(\cos E-e + \i \sqrt{1-e^2} \sin E\right)
    \expo{-\i k (E-e\sin E)} \d E.
  \end{equation}
  From this last expression, one only has to expand $\sqrt{1-e^2}$ and $\expo{\i k e \sin E}$ in power series of $e$
  and integrate over $E$ to obtain the expansion of $X_k$.
  Equations~(\ref{eq:X1p})-~(\ref{eq:X2m}) provide this expansion for the four coefficients of interest in this study
  ($X_{\pm1}$, $X_{\pm2}$).

  Equation~(\ref{eq:hansenE}) is also useful to numerically estimate $X_k$.
  One simply needs to sample $E$ over $[0,2\pi]$ and compute the integral (rectangle method) of Eq.~(\ref{eq:hansenE}):
  \begin{equation}
    \label{eq:numHansen}
    X_k \approx \frac{1}{N} \sum_{k=1}^{N}  \left(\cos E_k-e + \i \sqrt{1-e^2} \sin E_k\right)
    \expo{-\i k (E_k-e\sin E_k)},
  \end{equation}
  with $E_k = 2\pi \frac{k-1}{N}$.
  \figAI
  Figure~\ref{fig:AI} illustrates how the numerical estimates of Eq.~(\ref{eq:numHansen})
  evolve as a function of $N$ (for $X_1(0,99)$).
  We observe that when $N$ is too small, the rectangle method is not precise enough
  and when $N$ is too large, machine errors accumulate and the precision is lost.
  In this study, we use $N=100$, which seems to be a good compromise (we tested different Hansen coefficients
  and different values of $e$).

  \section{Derivatives of Hansen coefficients}
  \label{sec:DerivHansenCoeff}

  The computation of derivatives of Hansen coefficients is very similar to the computation of the coefficients themselves
  (see Appendix~\ref{sec:CompHansenCoeff}).
  From Eq.~(\ref{eq:hansenE}) we deduce
  \begin{equation}
    \begin{array}{ll}
      \displaystyle\frac{\d X_k}{\d e} = \frac{1}{2\pi} \int_0^{2\pi} &\Bigg[
      \i k \sin E \left(\cos E-e + \i \sqrt{1-e^2} \sin E\right)\\
      &-\left(1 + \i \frac{e}{\sqrt{1-e^2}} \sin E\right) \Bigg]
      \expo{-\i k (E-e\sin E)} \d E.
      \label{eq:dhansenE}
    \end{array}
  \end{equation}
  As explained in Appendix~\ref{sec:CompHansenCoeff}, this expression can be used
  both to expand the derivatives in power series of $e$, and to numerically
  estimate the integral (rectangle method).

  \section{Orbital elements from extrema}
  \label{sec:extrema}

  In some cases, the Fourier coefficients cannot be  determined correctly (e.g. owing to a bad sampling).
  An alternative method, using extrema of the radial velocity curve, can be used to obtain a rough
  estimate of orbital parameters.
  We assume that the orbital period (as well as instrumental offsets and drifts)
  has been correctly determined but not the Fourier coefficients.
  By looking at the timings and values of the maximum and minimum of the folded radial velocity curve,
  we again obtain  four observables that allow  the four parameters to be determined:
  $K$, $e$, $\omega$, $M_0$.
  We note $V_{min}$, $V_{max}$, $t_{min}$, $t_{max}$ the values and timings (in $[0,P]$) of the minimum
  and maximum, assuming that the mean value of $V$ is zero.
  From Eq.~(\ref{eq:rv1}), we obtain
  \begin{eqnarray}
    \label{eq:Vmin} V_{min} &=& K (e\cos\omega - 1),\\
    \label{eq:Vmax} V_{max} &=& K (e\cos\omega + 1),\\
    \label{eq:vmin} v_{min} &=& \pi - \omega,\\
    \label{eq:vmax} v_{max} &=& - \omega.
  \end{eqnarray}
  We have (see Eqs.~(\ref{eq:Vmin}),~(\ref{eq:Vmax}))
  \begin{eqnarray}
    \label{eq:estKminmax}
    \hat{K} &=& \frac{V_{max}-V_{min}}{2},\\
    \label{eq:estecwminmax}
    \hat{e}\cos\hat{\omega} &=& \frac{V_{max}+V_{min}}{V_{max}-V_{min}}.
  \end{eqnarray}
  To determine the remaining parameters ($e\sin\omega$, $M_0$),
  we need to use the information contained in the timing of extrema.
  We have
  \begin{equation}
    M = M_0 +  n t,
  \end{equation}
  thus
  \begin{eqnarray}
    \label{eq:Mmin}
    M_{min} &=& M_0 + n t_{min},\\
    \label{eq:Mmax}
    M_{max} &=& M_0 + n t_{max}.
  \end{eqnarray}
  These mean anomalies ($M$) can be expressed as functions of the eccentricity and
  the true anomalies ($v$), which are given by Eqs.~(\ref{eq:vmin}),~(\ref{eq:vmax})
  \begin{eqnarray}
    \label{eq:ME}
    M &=& E - e\sin E,\\
    \label{eq:Ev}
    E &=& 2\arctan\left(\sqrt{\frac{1-e}{1+e}} \tan\left(\frac{v}{2}\right)\right).
  \end{eqnarray}
  Developing Eqs.~(\ref{eq:ME}),~(\ref{eq:Ev}) at second order in eccentricity, we obtain
  \begin{equation}
    M = v - 2 e \sin v + \frac{3}{4} e^2\sin 2 v + \O{e^3},
  \end{equation}
  which gives
  \begin{eqnarray}
    &&M_{min} = \pi - \omega - 2 e\sin\omega - \frac{3}{4} e^2 \sin 2\omega + \O{e^3},\\
    &&M_{max} = - \omega + 2 e\sin\omega - \frac{3}{4} e^2 \sin 2\omega + \O{e^3},\\
    &&M_{max}-M_{min} = \pi + 4 e\sin\omega + \O{e^3},
  \end{eqnarray}
  thus (see Eqs.~(\ref{eq:Mmin}),~(\ref{eq:Mmax}))
  \begin{equation}
    \label{eq:esteswminmax}
    \hat{e} \sin\hat{\omega} = \frac{n}{4}(t_{max}-t_{min}) - \frac{\pi}{4} \mod  \frac{\pi}{2}
  \end{equation}
  This provides an estimate for $e\sin\omega$, and thus for $e$ and $\omega$
  (since we already have an estimate for $e\cos\omega$ see Eq.~(\ref{eq:estecwminmax})).

  Finally, using the estimates for $e$ and $\omega$ in Eqs.~(\ref{eq:ME}),~(\ref{eq:Ev}),
  we obtain the values of $M_{min}$, $M_{max}$.
  The remaining parameter $M_0$ is straightforwardly derived from these values
  \begin{equation}
    \label{eq:estM0minmax}
    \hat{M}_0 = M_{min} - nt_{min} = M_{max} - nt_{max}.
  \end{equation}

  The estimates given in Eqs.~(\ref{eq:estKminmax}),~(\ref{eq:estecwminmax}),~(\ref{eq:esteswminmax}),~(\ref{eq:estM0minmax}) rely on the determination of the period $P$ and the values and timings of the extrema ($V_{min}$, $V_{max}$, $t_{min}$, and $t_{max}$).
  We use the period estimate given by the periodogram method (see Sect.~\ref{sec:periodo}).
  The extrema determination can be affected by offsets, drifts, and outliers.
  We thus substract from the data,
  the offsets and drifts determined by the linear fit used for the Fourier method (see Sect.~\ref{sec:periodo}).
  Moreover, to reduce the impact of outliers, we compute the minimum and maximum (values and timings)
  as a weighted mean over the lowest and highest $N$ points
  (we chose $N=2$ for \object{HD~156846}b, see Sect.~\ref{sec:hd-156846}).
  This method provides very crude estimates for the parameters and is mainly useful when the Fourier method fails.

  \bibliographystyle{aa}
  \bibliography{DSBA}
\end{document}